\documentclass[a4paper]{spie}  
\usepackage{graphicx, subfigure}
\usepackage{float}
\usepackage{epsfig}
\usepackage{gensymb}
\usepackage{colortbl}
\usepackage{xcolor}
\usepackage{array}
\title{Computation and validation of two-dimensional PSF simulation based on physical optics}

\author{K.~Tayabaly\supit{1,2}, D.~Spiga\supit{1}, G.~Sironi\supit{1}, R.Canestrari\supit{1}, M.Lavagna\supit{2}, G.~Pareschi\supit{1}
\skiplinehalf
\supit{1} INAF/Brera Astronomical Observatory, Via Bianchi 46, 23807 Merate, Italy\\
\supit{2} Politecnico di Milano, Via La Masa 1, 20156 Milano, Italy}

\authorinfo{contact author: E-mail: \underline{kashmira.tayabaly@brera.inaf.it}, Telephone: +39 02 72320428}

\begin{document} 
\maketitle 

\begin{abstract}

The Point Spread Function (PSF) is a key figure of merit for specifying the angular resolution of optical systems and, as the demand for higher and higher angular resolution increases, the problem of surface finishing must be taken seriously even in optical telescopes. From the optical design of the instrument, reliable ray-tracing routines allow computing and display of the PSF based on geometrical optics. However, such an approach does not directly account for the scattering caused by surface microroughness, which is interferential in nature. Although the scattering effect can be separately modeled, its inclusion in the ray-tracing routine requires assumptions that are difficult to verify. In that context, a purely physical optics approach is more appropriate as it remains valid regardless of the shape and size of the defects appearing on the optical surface. Such a computation, when performed in two-dimensional consideration, is memory and time consuming because it requires one to process a surface map with a few micron resolution, and the situation becomes even more complicated in case of optical systems characterized by more than one reflection. Fortunately, the computation is significantly simplified in far-field configuration, since the computation involves only a sequence of Fourier Transforms. In this paper, we provide validation of the PSF simulation with Physical Optics approach through comparison with real PSF measurement data in the case of ASTRI-SST M1 hexagonal segments. These results represent a first foundation stone for future development in a more advanced computation taking into account microroughness and multiple reflection in optical systems.

\end{abstract}


\section{INTRODUCTION}
\label{sec:intro} 
The Point Spread Function (PSF) is a key figure of merit to characterize optical systems, and it is highly dependent on the optical design, quality of optical components and their alignment. While computing the optical performances of an optical system, one has to simulate the real PSF including fabrication errors such as shaping errors and surface finishing. Regarding the former, well-established ray-tracing routines allow computing and display of the PSF based on geometrical optics\cite{Sironi2014} but do not directly account for the scattering caused by surface microroughness. In fact, scattering is a typical interferential effect and cannot be described via geometrical methods. Although the scattering effect can be separately modeled, its inclusion in the ray-tracing routine requires assumptions (e.g., that the scattering and the geometry work independently from each other; that mid-frequency errors can be treated with geometric methods) that are difficult to verify. Therefore, adopting a purely physical optics approach is safer, because its validity is unrestricted. More exactly, it can be applied to any defect appearing on the optical surface, incidence angle, and light wavelength. 

The problem of computing the PSF of a focusing mirror including the roughness effect has been treated extensively in\cite{RaiSpi15, SpiRai14} for the case of grazing incidence mirrors. In that approach, the computation of the Fresnel integral in single or double reflection was simplified because the effects of mirror imperfections in grazing incidence are concentrated in the incidence plane, and also the impact of profile errors in the sagittal plane is usually negligible. For near-normal incidence mirrors, such as Cherenkov telescopes, these approximations are no longer applicable and the formalism has to forcedly be two-dimensional. Unfortunately, the numerical computation of Fresnel integrals in 2D leads to an unbearable computational complexity if the mirror surface is sampled at a few micron step. As a possible way out, however, we can limit ourselves to the case of a PSF evaluated at a distance that is much larger than the linear size of the mirror (far-field approximation). In these conditions, the PSF computation is sped up since the expression for the diffracted electric field reduces to the Fourier Transform of the Complex Pupil Function (CPF). This result was also found when dealing with the far-field PSF for grazing incidence mirrors\cite{RaiSpi15}, and can be applied to mirrors of any profile, characterized by any kind of surface error, and to any light wavelength. In this paper, we put in place the foundation stone for generalizing this approach to the two-dimensional case. Indeed, on the basis of physical optics, we theoretically derive the PSF after a single reflection and prove the validity of the corresponding off-axis PSF computation (PO) by comparing the results to Ray Tracing (RT) approach and experimental data for the aspherical ASTRI-SST M1 hexagonal segments. ASTRI-SST is a prototype of a Small Sized Telescope (SST) for the international CTA \cite{DesignCTA} project dedicated to the study of gamma ray sources. This prototype was entirely designed, manufactured and integrated by INAF-OAB\cite{Canestrari14} who adopted a Schwarzschild-Couder configuration. 

For optical telescopes' applications, the formalism has also to be extended to multiple reflections taking into account a source that could be off-axis and not at infinite distance such as for ASTRI-SST \cite{Pareschi14}. Such a computation will be described in a subsequent work. 

\section{Single-reflection optical systems: Point Spread Function computation in far-field approximation}
\label{sec:PSF_POtheo}

\begin{figure}[hbt]
	\centering
     \includegraphics[width = 0.6\textwidth]{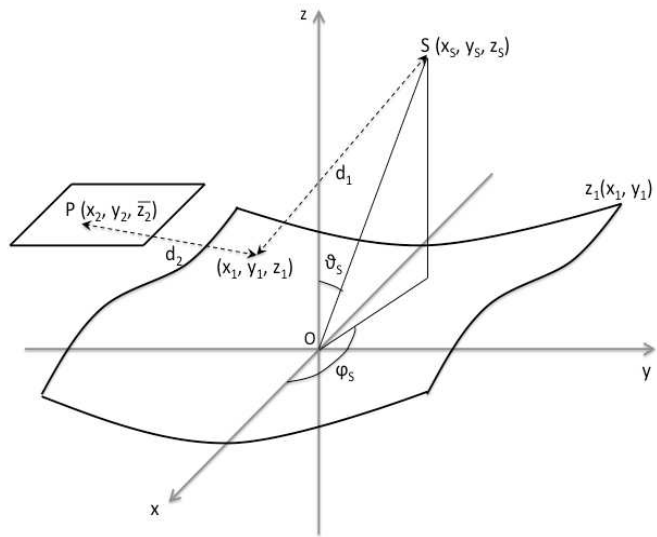}
	\caption{Scheme of computation for the electric field from the source S, diffracted by a mirror described by the profile $z_1$($x_1$, $y_1$) at the ($x_2$, $y_2$, $\bar{z_2}$) location on the detection plane (or the secondary mirror surface). Polarization effects are not considered. The polar angles that locate the position of the detector are omitted to simplify the drawing.}
	\label{fig:scheme}
\end{figure}

Following the procedure based on the Huygens-Fresnel principle and developed \cite{RaiSpi15} for the case of grazing-incidence mirrors, we adopt a reference frame as in Fig.~\ref{fig:scheme}, with the mirror profile described by the function $z_1$($x_1$, $y_1$), over a pupil M$_1$ of generic shape and size. For example, for the ASTRI panels, M$_1$ is hexagonal. We select the origin of the reference frame setting $z_1$(0,0)=0. The source S, located at ($x_{\mathrm S}$, $y_{\mathrm S}$, $z_{\mathrm S}$), is assumed to be point-like and perfectly monochromatic of wavelength $\lambda$; hence, spatially and temporally coherent. The diffracted intensity is recorded at the position P($x_2$, $y_2$, $\bar{z_2}$), which may represent a location of either a detector array or the surface of the secondary mirror, if included in the optical layout. The detection array is parallel to the $xy$ plane, at a constant height $\bar{z_2}$. We now assume that $\bar{z_2} \gg z_1$, $z_{\mathrm S} \gg z_1$ over the entire mirror map, and finally that $\bar{z_2}$ and $z_{\mathrm S}$ are much larger than the mirror lateral size. This approximation will enable us to reduce the Fresnel integrals to the Fourier Transform.

As per the Fresnel diffraction theory, the electric field at ($x_2$, $y_2$, $\bar{z_2}$) is 
\begin{equation}
	E_2(x,y, \bar{z_2}) = \int_{\mathrm{M_1}}\frac{E_0 q_{12}}{\lambda d_2} \,e^{-\frac{2\pi {\mathrm i}}{\lambda}(d_1+d_2)}\,\mbox{d}r^2_{\perp}, 
	\label{eq:fres}
\end{equation}
where $E_0$ is the electric field amplitude at the mirror surface, assumed to be constant. We also assume the amplitude reflectivity to be constant everywhere, in order to take it out of the integral sign; to simplify the notation, we assume it equal to 1. The obliquity factor $q_{12}$ depends on the angle between the scattered and the incident directions and, in near-normal geometry, can be assumed to be approximately 1. The orthogonal surface element term can be written as $\mbox{d}x_1 \mbox{d}y_1 \cos\theta_2$, where $\theta_2$ is the angle between the surface normal and the $d_2$ direction. This angle is usually assumed to be constant if the mirror curvature is not too pronounced. Finally, we denote with $d_1$ and $d_2$ the distances of the generic mirror point from S and P, respectively:
\begin{eqnarray}
d_1 & = & \sqrt{(x_{\mathrm S}-x_1)^2+(y_{\mathrm S}-y_1)^2+(z_{\mathrm S}-z_1)^2},\label{eq:d1} \\
d_2 & = & \sqrt{(x_2-x_1)^2+(y_2-y_1)^2+(\bar{z_2}-z_1)^2}. \label{eq:d2} 
\end{eqnarray}
At a large distance from the mirror, we can approximate the roots in the two previous equations at the first order:
\begin{eqnarray}
d_1 & \simeq & (z_{\mathrm S}-z_1)+\frac{(x_{\mathrm S}-x_1)^2+(y_{\mathrm S}-y_1)^2}{2(z_{\mathrm S}-z_1)}, \label{eq:d1_app} \\
d_2 & \simeq & (\bar{z_2}-z_1)+\frac{(x_2-x_1)^2+(y_2-y_1)^2}{2(\bar{z_2}-z_1)}. \label{eq:d2_app} 
\end{eqnarray}
This definition of $d_1$ diverges to infinity in the case of a source at infinite distance. To remove this singularity, we arbitrarily set the reference wavefront passing by the reference frame origin. Approximating the denominators with $z_{\mathrm S}$ and $\bar{z_2}$, and removing unessential phase constants, we remain with
\begin{eqnarray}
d_1 & \simeq & -z_1+\frac{x_1^2-2x_{\mathrm S}x_1+y_1^2-2y_{\mathrm S}y_1}{2z_{\mathrm S}}, \label{eq:dis1} \\
d_2 & \simeq & -z_1+\frac{x_1^2-2x_2x_1+y_1^2-2y_2y_1}{2 \bar{z_2}}+\frac{x_2^2+y_2^2}{2\bar{z_2}}. \label{eq:dis2} 
\end{eqnarray}
In the last equation, we have maintained and isolated the dependence on $x_2$ and $y_2$ to preserve the information on the phase of the diffracted wavefront. This will enable diffraction calculation in a future development of the work. Substituting the expressions we obtain:
\begin{equation}
	E_2(x, y, \bar{z_2}) = \frac{E_0}{\lambda \bar{z_2}}\cos\theta_2\,e^{-\frac{\pi {\mathrm i}}{\lambda \bar{z_2}}(x_2^2+y_2^2)}\int_{\mathrm{M_1}} e^{\frac{2\pi {\mathrm i}}{\lambda}2\left(z_1-\frac{x_1^2+y_1^2}{4z_{\mathrm R}}\right)} e^{2\pi {\mathrm i}(f_xx_1+f_yy_1)} \,\mbox{d}x_1\mbox{d}y_1, 
	\label{eq:E_diff}
\end{equation}
where we have defined 
\begin{equation}
	\frac{1}{z_{\mathrm R}} = \frac{1}{z_{\mathrm S}}+\frac{1}{\bar{z_2}}, 
	\label{eq:zR}
\end{equation}
and the spatial frequencies via the equations:
\begin{equation}
	f_x = \frac{1}{\lambda}\left(\frac{x_2}{\bar{z_2}}+\frac{x_{\mathrm S}}{z_{\mathrm S}}\right), \hspace{2cm} f_y = \frac{1}{\lambda}\left(\frac{y_2}{\bar{z_2}}+\frac{y_{\mathrm S}}{z_{\mathrm S}}\right).
	\label{eq:sp_freq}
\end{equation}
The specular direction of the source is correctly defined as $x_2/\bar{z_2} = - x_{\mathrm S}/z_{\mathrm S}$, and similarly for $y_2$. We notice that the case of a source at infinity is easy to treat, replacing the coordinate ratios of the source with
\begin{equation}
	f_x^{\infty} = \frac{1}{\lambda}\left(\frac{x_2}{\bar{z_2}}+\sin\theta_{\mathrm S}\cos{\varphi_{\mathrm S}}\right), \hspace{2cm} f_y^{\infty} = \frac{1}{\lambda}\left(\frac{y_2}{\bar{z_2}}+\sin\theta_{\mathrm S}\sin{\varphi_{\mathrm S}}\right),
	\label{eq:sp_freq_inf}
\end{equation}
where $\theta_{\mathrm S}$ and $\varphi_{\mathrm S}$ are the zenithal and the azimuthal angle of the source, with respect to the origin of the reference frame (see Fig.~\ref{fig:scheme}). 

The result above shows that the electric field at $\bar{z_2}$ is the Fourier Transform of the Complex Pupil Function,
\begin{equation}
	\mbox{CPF}(x_1, y_1) = \chi_{\mathrm{M_1}}(x_1, y_1) \, e^{\frac{4\pi {\mathrm i}}{\lambda}\Delta z_1} ,
	\label{eq:CPF}
\end{equation}
where $\chi_{\mathrm{M1}}$ is the characteristic function of the pupil, and $\Delta z_1$ is defined as:
\begin{equation}
	\Delta z_1 = z_1(x_1, y_1)-\frac{x_1^2+y_1^2}{4z_{\mathrm R}}, 
	\label{eq:Dz}
\end{equation}
i.e., the difference between the true mirror profile and a parabola with focal length $z_{\mathrm R}$, which acts as a reference surface. 

We note that the expression of the reference surface does not include $x_{\mathrm S}$. This is quite surprising because one would expect to have as a reference surface the profile that exactly images the source into the center of the detector plane. For example, for a source at finite distance the surface to be removed from $z_1$ would expectedly be an ellipse, as done in grazing incidence mirrors under the same approximations\cite{RaiSpi15}. However, in grazing incidence the mirror defect direction is essentially parallel to the coordinate on the detector. Therefore, in far-field approximation executing a Fourier Transform of the profile errors is the natural choice. In near-normal incidence, in contrast, the mirror profile error $\Delta z$ is {\it orthogonal} to the $x_2, y_2$ detector coordinates: hence, we may expect a change in the reference surface when we perform the Fourier Transform of the CPF.

The intensity on the detector plane is represented by the square module of the electric field at $\bar{z_2}$:
\begin{equation}
	I(x_2, y_2, \bar{z_2}) =\left|E_2(x_2, y_2, \bar{z_2})\right|^2, 
	\label{eq:flux}
\end{equation}
and the power intercepted by the mirror from the source can be written as:
\begin{equation}
	I_0 = E_0^2 \cos\theta_{\mathrm S}A_{{\mathrm M_1},\perp}, 
	\label{eq:I_0}
\end{equation}
where $A_{{\mathrm M_1},\perp} = A_{\mathrm M_1}\cos\theta_{\mathrm S}$ is the projection of the pupil area in the direction orthogonal to the incidence direction. The additional factor of $\cos\theta_{\mathrm S}$ in Eq.~\ref{eq:I_0} accounts for the intensity spread as per Lambert law. We obtain the PSF normalizing the Eq.~\ref{eq:flux} to $I_0$, and substituting the expression of $E_2(x_2, y_2, \bar{z_2})$ (Eq.~\ref{eq:E_diff}) we obtain:
\begin{equation}
\mbox{PSF}(x_2, y_2, \bar{z_2}) = \frac{1}{\lambda^2 \bar{z_2}^2A_{\mathrm M_1}}\left|\int_{\mathrm P1} e^{\frac{2\pi {\mathrm i}}{\lambda}2\Delta z_1} e^{2\pi {\mathrm i}(f_xx_1+f_yy_1)} \,\mbox{d}x_1\mbox{d}y_1\right|^2, 
	\label{eq:PSF}
\end{equation}
where we have assumed that the PSF is observed in the vicinities of the specular direction, and therefore $\theta_{\mathrm S} \simeq \theta_2$. The multiplicative factor ensures that the integrated PSF is normalized to 1:
\begin{equation}
	\int_{-\infty}^{+\infty}\!\mbox{d}y_2 \int_{-\infty}^{+\infty}\! \mbox{PSF}(x_2,y_2) \,\mbox{d}x_2= 1.
	\label{eq:PSFnorm}
\end{equation}

Except for a few cases, the Fourier Transform in Eq.~\ref{eq:PSF} has to be computed numerically. To this end, a proper spatial sampling of the pupil, $\Delta x_1 $, is requested to avoid aliasing in the PSF within a detector field of lateral size $L_2$. This is obtained by a straightforward extension of the monodimensional case\cite{RaiSpi15},
\begin{equation}
	\Delta x_1 = \frac{\lambda \bar{z_2}}{2\pi L_2\cos\theta_{\mathrm S}}.
	\label{eq:Dx}
\end{equation}
The mirror pupil can have any shape, but it should always be contained in a square of side twice as large as the maximum diameter of the pupil, $L_1$. The $z_1$ map is therefore a $N\times N$ square matrix, with 
\begin{equation}
	N = \frac{4\pi L_2L_1\cos\theta_{\mathrm S}}{\lambda \bar{z_2}}.
	\label{eq:N}
\end{equation}

In the next section, this PSF derivation is applied to the particular case of ASTRI-SST segmented primary mirror for validation. 

\section{Case Study: ASTRI-SST M1 hexagonal segments}
\subsection{ASTRI-SST M1 optical design}

\begin{figure}[h!]
\begin{center}
\begin{tabular}{p{8cm} p{8cm}}
\centering {\includegraphics[height = 8cm,width=8cm]{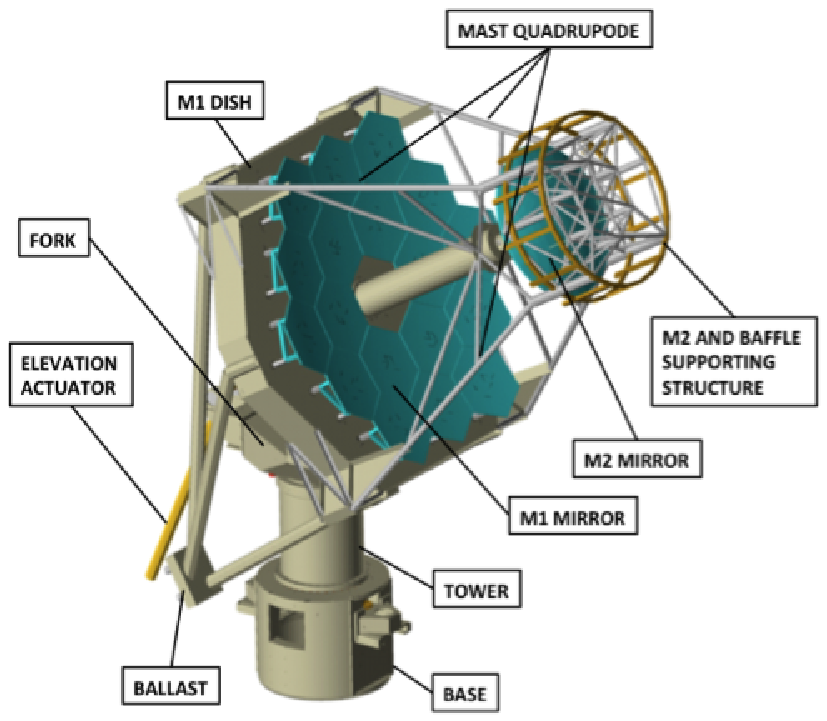}} & {\includegraphics[height = 7cm,width=8cm]{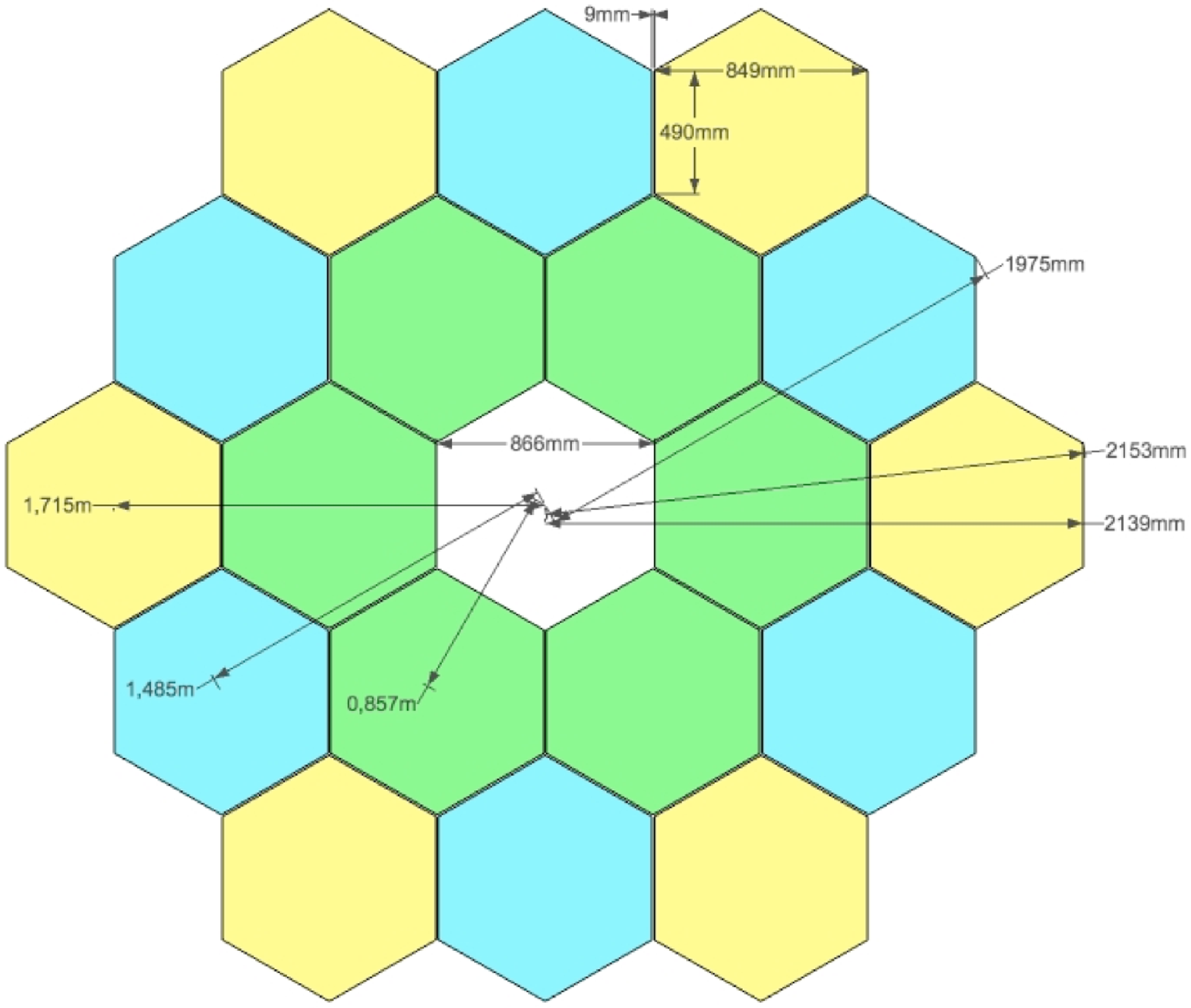}}\\ 
\centering A)  & \centering B) \\
\end{tabular}
\caption{Optical design of ASTRI-SST (after\cite{Canestrari14}). A) 3D optical layout for ASTRI-SST 2M telescope. B) ASTRI-SST M1 tessellation. Green: corona 1. Blue: corona 2. Yellow: corona 3. \label{fig:ASTRIDesign}}
\end{center}
\end{figure}

ASTRI (Fig.~\ref{fig:ASTRIDesign}, A) is a prototype for CTA-SST fully developed by INAF-OAB \cite{Canestrari14}. Its optical design is based on a Schwarzschild-Couder configuration with a F\# of 0.5 and an equivalent focal length of 2150~mm. The M1 mirror diameter is 4.02~m in diameter, tessellated into hexagonal mirrors which face-to-face dimension is 849~mm. The M1 mirror is divided into 3 sections or 3 coronae, according to the distance of the center of the segment to the center of M1 (Fig.~\ref{fig:ASTRIDesign}, B). Therefore, the closest segments (857~mm from the center of M1) belong to corona 1, mirrors placed 1485~mm from the center of M1 are in corona 2, and the farthest segments (1715~mm from the center of M1) are in corona 3.

The M1 mirror of ASTRI-SST is an aspherical mirror defined as: 
\begin{equation}
z(r) = \frac{cr^2}{{1+\sqrt{1-(cr)^2}}} + \sum_{i=1}^{N} {a_i r^{2i}}
	\label{eq:M1geom}
\end{equation}
where $z(r)$ is the surface profile, $r$ the surface radial coordinate, $c$ the curvature and $\alpha_i$ the coefficients of asphericity detailed in Tab.~\ref{tab: CoeffAsphM1ASTRI}. 
	
	\begin{table}[h]
	\centering
	\begin{tabular}{|c|c|}
	\hline
	\textbf{i }& \textbf{$\alpha$}\\ 
	\hline
	\textbf{1} & 0\\ 
	\textbf{2}& $9.61060 *10^{-13}$\\ 
	\textbf{3}& $-5.65501 10^{-20}$\\ 
	\textbf{4}& $6.77984 10^{-27}$\\ 
	\textbf{5}& $3.89558 10^{-33}$\\ 
	\textbf{6}& $5.28038 10^{-40}$\\ 
	\textbf{7}& $-2.99107 10^{-47}$\\ 
	\textbf{8}& $-4.39153*10^{-53}$\\ 
	\textbf{9}& $-6.17433 10^{-60}$\\ 
	\textbf{10}& $2.73586*10^{-66}$\\
	\hline
	\end{tabular}
	\vspace{2mm}
    \caption{Coefficients of asphericity for ASTRI-SST M1 mirror}
	\label{tab: CoeffAsphM1ASTRI}
	\end{table}

\subsection{Experimental setup for PSF acquisition}

The experimental PSF is acquired using a setup developed at INAF-OAB and described in details in\cite{Sironi2014}. The setup consists of:

\begin{itemize}
\item \textbf{An optical bench} hosting the light source at 8430~mm from the center of the mirror. 
\item \textbf{A mirror support} composed of  a mechanical support on top of a rotary table where the ASTRI-SST M1 panels are mounted, and where tip and tilt could also be adjusted. The rotary table allows the rotation of the mirror toward the optical bench and the screen. For our study, an hexagonal segment from corona 3 area of ASTRI-SST M1 (Fig.~\ref{fig:ASTRIDesign},B) was mounted and oriented at 8.79~deg angle between the normal of the mirror at its center and the axis passing by the light source and the center of the mirror. 
\item \textbf{A screen} is mounted on a stage that allows moving the screen at different distances from the mirror. PSF acquisition have been performed placing the screen at 4761~mm, 5794~mm, 7059~mm, 8183~mm, and 8840~mm from the mirror. The stage direction is aligned with the center of the rotary table at the basis of the mirror support by using a laser. The center of the rotary table is positioned on the barycenter of the obtained pattern. 
\end{itemize}

\begin{figure}[hbt]
	\centering
     \includegraphics[width = 0.6\textwidth]{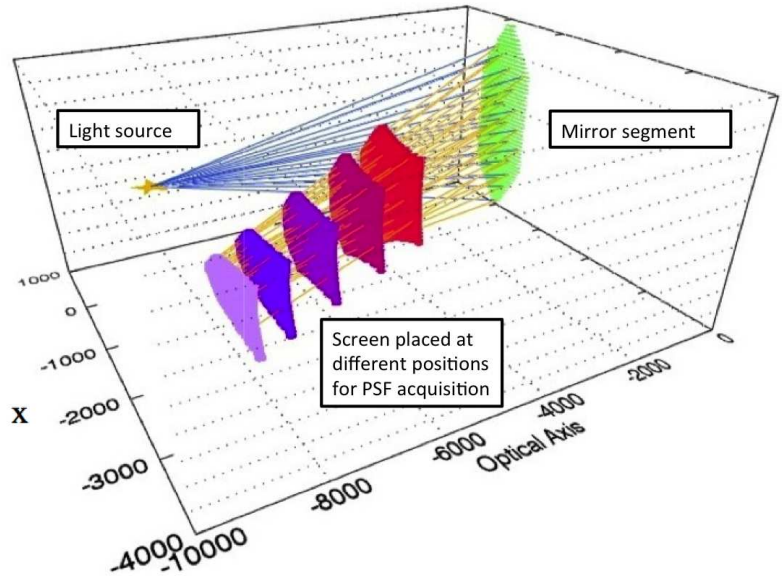}
\caption{Setup for the experimental PSF acquisition of ASTRI-SST corona 3 segment (after\cite{Sironi2014}): a light source is placed at 8430~mm from the center of the mirror panel that is rotated at 8.79~deg off-normal, the reflected image generated by the mirror is acquired at various distances from the center of the mirror (4761~mm, 5794~mm, 7059~mm, 8183~mm, and 8840~mm)} 
	\label{fig:ExperimentalPSFsetup}
\end{figure}

The facility uses a folding mirror to acquire distances perpendicularly to the optical axis and a camera is used to acquire pictures of the images reflected on the screen. Fig.~\ref{fig:ExperimentalPSFsetup} shows the setup configuration in 3D for experimental PSF acquisition. 

\subsection{Comparing PSF obtained from physical optics (PO), ray tracing (RT) and experimental data}
\subsubsection{Physical Optics}

\begin{figure}[h] 
  \begin{center}
  \begin{tabular}{p{8cm}p{8cm}}
 {\includegraphics[height = 8cm,width=7.3cm]{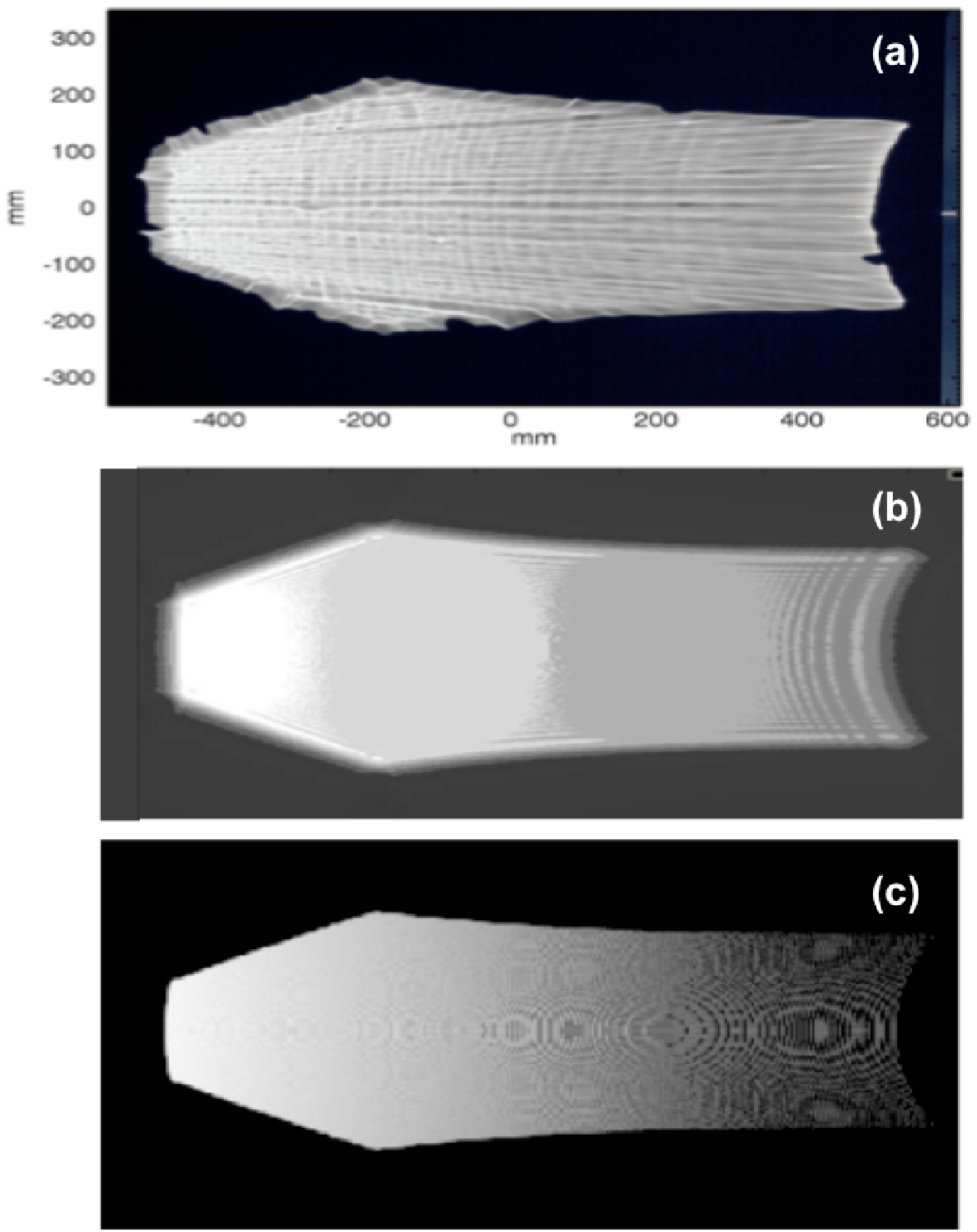}} &{\includegraphics[height = 8cm,width=7.3cm]{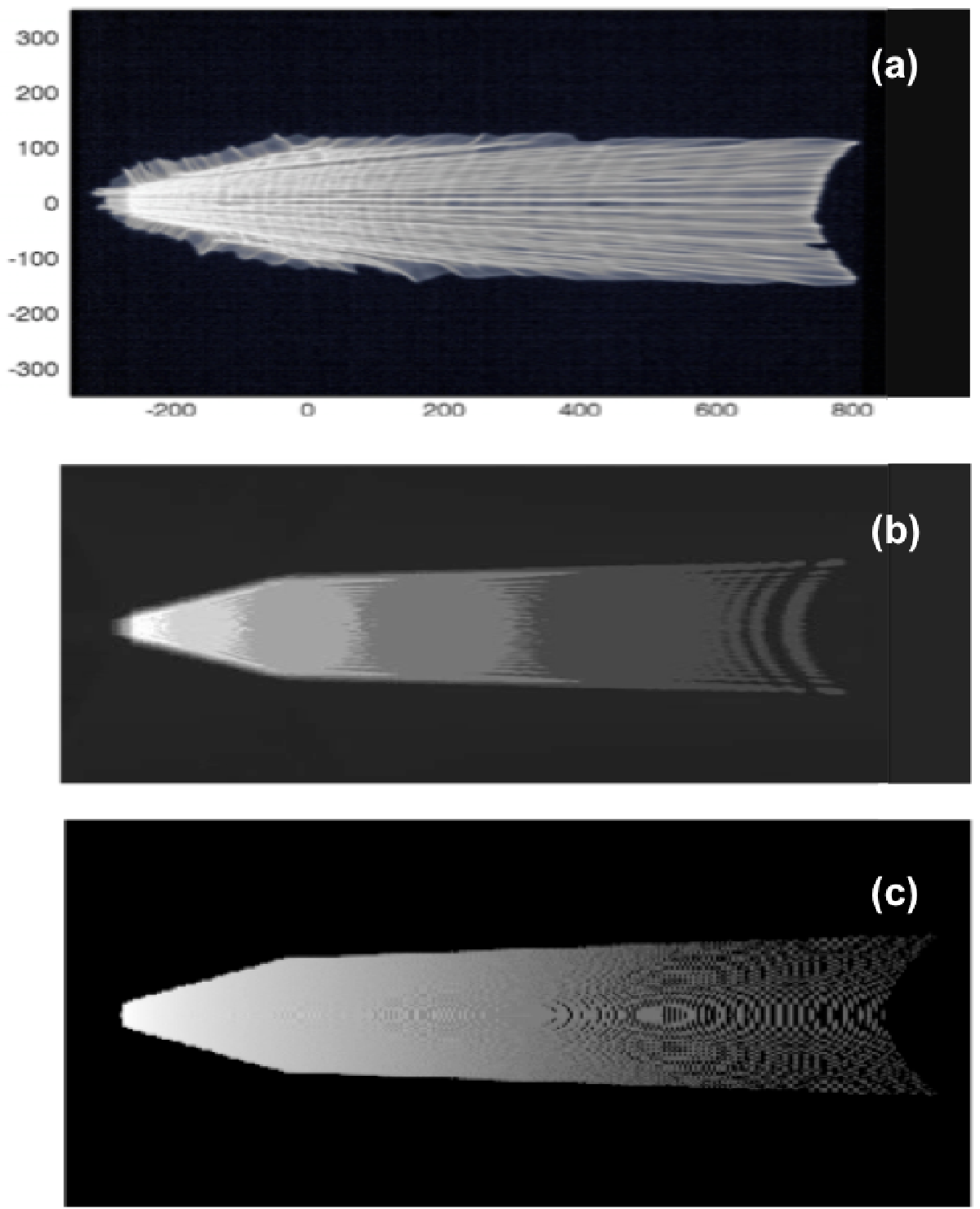}}\\ \centering
\small{$d_{\mbox{DM}}$= 5794~mm } & \centering \small{$d_{\mbox{DM}}$= 8183~mm}\\
\end{tabular}
 \caption{Comparison of PSF images on the same scale obtained from (a) Experimental image, (b) Physical Optics computation at $\lambda$ = 100~$\mu$m and (c) Ray Tracing computation. The mirror was oriented at 8.79~deg, from a light source located at 8430~mm from the center of ASTRI-SST M1 corona 3 mirror segment and a varying distance of the detector $d_{\mbox{DM}}$}
\label{fig: PSFimagecompa}
\end{center}
\end{figure}

The PSF obtained from Physical Optics (PO) approach is computed via a Matlab code that applies the theory expounded in Sect.~\ref{sec:PSF_POtheo}. In order to validate this approach, the results are compared to the Ray Tracing program TraceIT\cite{Sironi2011} and to experimental data obtained for ASTRI-SST M1 corona 3 segment as illustrated in Fig.~\ref{fig: PSFimagecompa}. To decrease the computation time, the PO simulation considers the incoming light at $\lambda$ = 100~$\mu$m which is much higher than the light source used in the experience. This does not change the shape or the size of the PSF but introduces diffraction fringes clearly noticeable in Fig.~\ref{fig: PSFimagecompa}(b). Thence not only does the PO approach perfectly reproduces the Ray Tracing (or geometrical optics) PSF prediction but takes also in consideration diffraction effect in the computation that manifests itself as diffraction fringes in the images shown here. Obviously, those fringes will be less extended and narrower for shorter wavelengths. 

\subsubsection{Comparing different approaches: PO, RT and experimental data}

\begin{figure}[h]
	\centering
     \includegraphics[width = 0.5\textwidth]{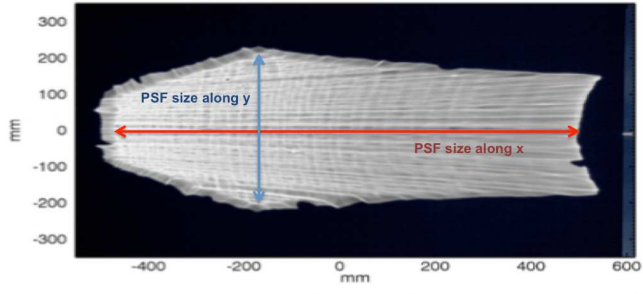}
\caption{Definition of PSF sizes along x and y} 
	\label{fig: PSFDimension}
\end{figure}

\parbox{\textwidth} {In order to quantify the agreement between the three PSFs obtained using three independent methods, we compare the size of the PSF in both $x$ and $y$ direction. At the edges, the experimental PSF is blurred, mainly due to defects on the mirror, this introduces uncertainty on the real PSF size. This uncertainty is estimated at 5\% and 10\% respectively in $x$ and $y$ direction. As the defects on the mirrors, since geometrical defects or roughness are not taken into account in the PO or RT simulations shown in this paper, the dimension of the experimental PSF appearing in Tab.~\ref{tab: PSFtableX} and~\ref{tab: PSFtableY} do not take into account the blurs at the edges of the PSF as shown in Fig.~\ref{fig: PSFDimension}.}

\begin{figure}[hbt]
	\centering
     \includegraphics[width = 0.8\textwidth]{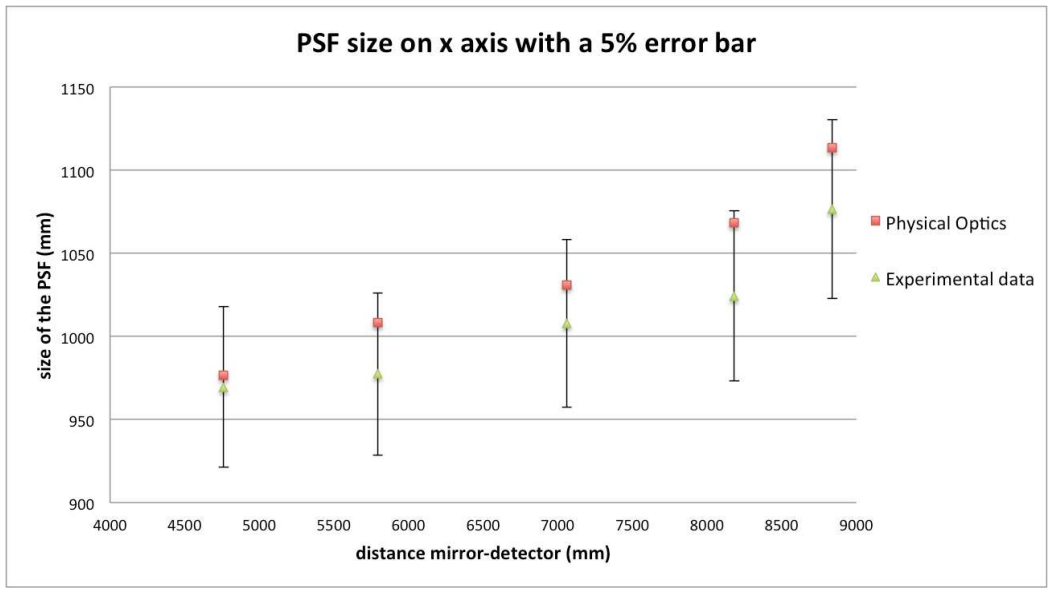}
\caption{Comparison of the PSF size obtained using Physical Optics (red squares) and Experimental images (green triangles) for a mirror oriented at 8.79~deg from the source located at 8430~mm from the center of the mirror with a $\pm$~5\% error bar on experimental data} 
	\label{fig:PSFCompaX}
\end{figure}

\begin{figure}[hbt]
	\centering
     \includegraphics[width = 0.8\textwidth]{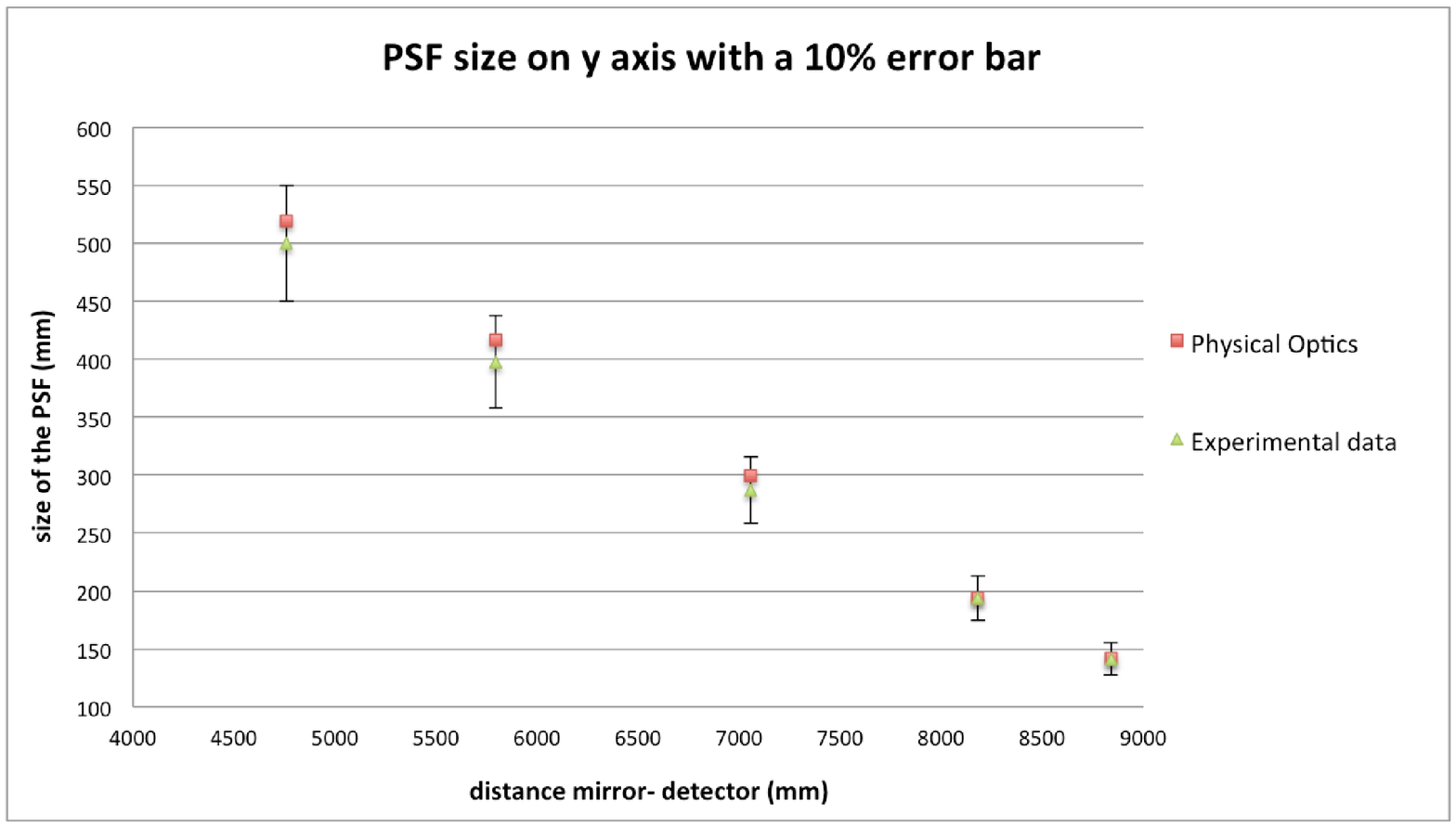}
\caption{Comparison of the PSF size along y obtained using Physical Optics (red squares) and Experimental images (green triangles) for a mirror oriented at 8.79~deg from the source located at 8430~mm from the center of the mirror with a $\pm$~10\% error bar on experimental data} 
	\label{fig:PSFCompaY}
\end{figure}

\parbox{\textwidth}{Hence, Fig.~\ref{fig:PSFCompaX} and~\ref{fig:PSFCompaY} show very good agreement between the PO simulation and the experimental data. For each considered case, the size of the PSF found using PO remains within the existing size uncertainty for experimental data. Moreover, although Ray Tracing data are slightly overestimating the size of the PSFs in the $x$ direction, given the various sources of error (experimental setup, PSF measurement accuracy, different simulation approaches, ...), the differences between each of the method is negligible as one can see in Tab.~\ref{tab: PSFtableX} and~\ref{tab: PSFtableY}. The prediction between PO, RT and real experimental data give self-consistent results giving confidence in the validity of the Physical Optics computation approach for single reflection. }

\vspace{5mm}

	\begin{table} [hbt]
	\centering
	\begin{tabular}{|>{\centering\arraybackslash}m{.1\textwidth} |>{\centering\arraybackslash}m{.1\textwidth}| >{\centering\arraybackslash}m{.1\textwidth}|>{\centering\arraybackslash}m{.1\textwidth}|>{\centering\arraybackslash}m{.1\textwidth}|>{\centering\arraybackslash}m{.1\textwidth}|} 
	\hline
	\rowcolor{gray} \textcolor{white}{\textbf{Distance} }&  \textcolor{white}{\textbf{RT (mm)}} &  \textcolor{white}{\textbf{PO (mm)}}& \textcolor{white}{\textbf{Exp (mm)}}& \textcolor{white}{\textbf{PO/Exp}} &\textcolor{white}{ \textbf{RT/Exp}} \\ 
	\hline
	\textcolor {black} {\textbf{4761}}& 1033.5 & 976.4 & 969.4 & \small \textbf{1.01} & \small \textbf{1.07}\\
	\hline 
	\textcolor {black} {\textbf{5784} }& 1061.5 & 1008.3 & 977.4 &  \small \textbf{1.03} & \small \textbf{1.09} \\ 
	\hline
	\textcolor {black} {\textbf{7059} }& 1097.2 & 1031.0 & 1007.8 &  \small \textbf{1.02} & \small \textbf{1.09} \\ 
	\hline
	\textcolor {black} {\textbf{8183} }& 1120.0 & 1068.5 & 1024.4 &  \small \textbf{1.04} &  \small \textbf{1.09} \\ 
	\hline
	\textcolor {black} {\textbf{8840} }& 1143.4 & 1113.7 & 1076.7 &  \small \textbf{1.03} & \small \textbf{1.06}\\ 
	\hline
	\end{tabular}
		\caption{PSF size along $x$ comparison.}
	\label{tab: PSFtableX}
	\end{table}

\vspace{5mm}
	\begin{table}[hbt]
	\centering
\begin{tabular}{|>{\centering\arraybackslash}m{.1\textwidth} |>{\centering\arraybackslash}m{.1\textwidth}| >{\centering\arraybackslash}m{.1\textwidth}|>{\centering\arraybackslash}m{.1\textwidth}|>{\centering\arraybackslash}m{.1\textwidth}|>{\centering\arraybackslash}m{.1\textwidth}|} 
	\hline
	\rowcolor{gray} \textcolor{white}{\textbf{Distance} }&  \textcolor{white}{\textbf{RT (mm)}} &  \textcolor{white}{\textbf{PO (mm)}}& \textcolor{white}{\textbf{Exp (mm)}}& \textcolor{white}{\textbf{PO/Exp}} &\textcolor{white}{ \textbf{RT/Exp}} \\ 
	\hline
	\centering
	\textbf{4761}& 524.2 & 518.8 & 500.0 & \small \textbf{1.04} & \small \textbf{1.05}\\
	\hline 
	\textbf{5784} & 426.4 & 416.2 & 397.5 &  \small \textbf{1.05} & \small \textbf{1.07} \\ 
	\hline
	\textbf{7059} & 304.3 & 299.6 & 287.1 &  \small \textbf{1.04} & \small \textbf{1.06}\\ 
	\hline
	\textbf{8183} & 207.5 & 194.0 & 194.0 &  \small \textbf{1.00} &  \small \textbf{1.07}\\ 
	\hline
	\textbf{8840} & 156.2 & 142.0 & 141.4 &  \small \textbf{1.01} & \small \textbf{1.10}\\ 
	\hline
	\end{tabular}
	\caption{PSF size along $y$ comparison.}
	\label{tab: PSFtableY}
	\end{table}

\section{Conclusion} 
\parbox{\textwidth}{In this paper, we have shown that a 2D PSF prediction, for mirrors working also close to normal incidence, based on Physical Optics considering a single reflection was feasible and accurate. This result is a foundation stone for the development of a complete tool to predict PSF with multi-reflection and mirror surface defects consideration in the study of scattering introduced by optical surfaces. The main advantage of such an approach is that it does not require any frequency regime separation consideration to study the scattering effects on a mirror. Indeed the different regimes are unified in the physical optics domain. Applied to X-ray optical systems at grazing incidence, this approach has already been successfully adopted in the 1D domain \cite{RaiSpi15} but has never been extended to the 2D domain. However, as we want to include surface defects of a few microns in meters scale optics, the resolution of matrices considered are strongly increased. A real computational challenge is rising, and it becomes even more daunting as we are considering complex systems (aspheric components and multiple reflection systems) for which the far-field approximation is not always verified. Solving those issues is currently under progress. }

\acknowledgments     

This work is supported by the Italian National Institute of Astrophysics (INAF) and the TECHE, T-REX programs funded by the Ministry of Education, University and Research (MIUR).

\bibliographystyle{spiebib}

\end{document}